\documentclass[11 pt]{article}
\usepackage{amsfonts,latexsym, amssymb, amsmath} 
\usepackage{graphicx,subfigure,epsfig} 

\newcounter{cst}

\newcounter{cexp}

\renewenvironment{abstract}{\begin{center} {\bf Abstract} \\
\end{center} }{\smallskip}

\usepackage{graphicx,amssymb,latexsym,amsmath,amsfonts,subfigure,epsf}
\usepackage[active]{srcltx}

\def\bfrho{\mbox{\boldmath$\rho$}}
\def\bfpsi{\mbox{\boldmath$\psi$}}
\def\bfthe{\mbox{\boldmath$\theta$}}

\begin{document}

\title{Turing Instability and Pattern Formation in an Activator-Inhibitor System with Nonlinear Diffusion}

\author{G. Gambino\footnote{Department of Mathematics, University of Palermo, Italy, gaetana@math.unipa.it}$\;$
M. C. Lombardo\footnote{Department of Mathematics, University of Palermo, Italy, lombardo@math.unipa.it}$\;$
M. Sammartino\footnote{Department of Mathematics, University of Palermo, Italy, marco@math.unipa.it} $\;$
}

\maketitle

\begin{abstract}
In this work we study the effect of density dependent nonlinear diffusion on pattern formation
in the Lengyel--Epstein system.
Via the linear stability analysis we determine both the Turing and the Hopf instability boundaries and we
show how nonlinear diffusion
intensifies the tendency to pattern formation;
in particular, unlike the case of classical linear  diffusion,
the Turing instability can occur even when diffusion of the inhibitor is significantly slower than
activator's one.
In the Turing pattern region we perform the WNL multiple
scales analysis to derive the equations
for the amplitude of the stationary pattern, both in the supercritical and in the
subcritical case.
Moreover, we compute the complex Ginzburg-Landau equation in the vicinity of the Hopf bifurcation point as it gives a slow spatio-temporal
modulation of the phase and amplitude of the
homogeneous oscillatory solution.

\end{abstract}
\vskip1cm
\begin{center}\textit{Accepted for publication in Acta Applicandae Mathematicae}\end{center}
\vskip1cm
\section{Introduction}
\label{intro}
%
%
Self-organized patterning in reaction-diffusion system driven by linear (Fickian) diffusion
has been extensively studied since the seminal paper of Turing. Nevertheless, in many experimental cases, the gradient of the density of one
species induces a flux of another species or of the species itself,
therefore nonlinear effects should be taken into account. Recently, nonlinear diffusion terms have appeared to model different
physical phenomena in diverse contexts like population dynamics and ecology
\cite{FCP13,MRW11,TLP10,GGJ03,Gal12,GV11,R-BT13,LRT14},
and chemical reactions \cite{BB12,KH11}.

The aim of this work is to describe the Turing
pattern formation for the following reaction-diffusion system with nonlinear density-dependent
diffusion:
\begin{equation}\label{or_syst}
\begin{split}
\displaystyle\frac{\partial U}{\partial \tau}&=D_u \displaystyle\frac{\partial}{\partial \zeta} \left(\left(\frac{U}{u_0}\right)^m\frac{\partial U}{\partial \zeta}\right)+\Gamma\left(a-U-\frac{4UV}{1+U^2}\right),\\
\displaystyle\frac{\partial V}{\partial \tau}&=c D_v \displaystyle\frac{\partial}{\partial \zeta} \left(\left(\frac{V}{v_0}\right)^n\frac{\partial V}{\partial \zeta}\right)+\Gamma c b\left(U-\frac{UV}{1+U^2}\right).
\end{split}
\end{equation}
In \eqref{or_syst}, $U(\zeta,\tau)$ and $V(\zeta,\tau)$, with $\zeta\in [0,\Omega], \Omega\in \mathbb{R}$,
represent the concentrations of two chemical species, the activator and
the inhibitor respectively; the reaction mechanism is chosen as in the Lengyel-Epstein system \cite{LE91,LE92} modeling
the chlorite-iodide-malonic acid and
starch (CIMA) reaction. The parameters $a$ and $b$ are positive constants related to the feed rate,
$c >1$ is a rescaling parameter which is bound up with starch concentration and
$\Gamma$ describes the relative strength of the reaction terms.

The nonlinear density-dependent diffusion terms, given by $D_u(U/u_0)^m$ and $D_v(V/v_0)^n$,
show that when $m,n>0$,
the species tend to diffuse faster (when $U>u_0$ and $V>v_0$) or slower (when $U<u_0$ and $V<v_0$)
than predicted by the linear diffusion. $D_u,D_v>0$ are the classical diffusion coefficients and
$u_0, v_0>0$ are threshold concentrations, measuring the strength of the interactions between the individuals of the same species.
\vskip 1cm
Nonlinear diffusion terms as in \eqref{or_syst}  could be employed to model autocatalytic chemical reactions occurring on porous media \cite{ZHMOL02}, or in networks of electrical circuits \cite{BPS07}, or on surfaces \cite{RW01}, like cellular membranes, or in surface electrodeposition \cite{BLMS12,BLS13}.

Various experimental and numerical studies have been conducted on the Lengyel--
Epstein system coupled with linear diffusion, see e.g. \cite{CDBD90,DCDB91}.
Also the analytical properties of the system
have been widely studied:
the Hopf bifurcation analysis has been performed in \cite{LM10};
Turing instability and the pattern formation driven by linear diffusion have been investigated in
different geometries \cite{YWS08,WWG12,WZ13,CK99}.
The existence and non-existence for the steady states of the system have been proved in \cite{NT05}.
To the best of our knowledge, the effect
of the nonlinear diffusivity on Turing pattern of the Lengyel--Epstein system has not
been examined, as exceptions we
mention \cite{YMH08}, where the authors determine the conditions for the occurring
of Turing instabilities when
linear diffusion for one species is coupled with the subdiffusion of the other species, and Ref. \cite{LHPLZS12}, where the authors perform an extensive numerical exploration of the Lengyel-Epstein
model with local concentration-dependent diffusivity.

In this paper we show that the nonlinear diffusion facilitate the Turing instability
and the formation of the Turing structures
as compared to the case of linear diffusion: in particular, increasing the value of the parameter $n$ in \eqref{or_syst}, the Turing instability
arises even when the diffusion of the inhibitor is significantly slower than
that of the activator (which is not the case when the diffusion is linear, i.e. when $n=0$, see \cite{NT05,WWG12}). Moreover, as the Lengyel-Epstein model also supports
the Hopf bifurcation, the formation of the Turing structure depends on the mutual location of the Hopf and Turing instability boundaries. Through  linear stability analysis we show that
increasing the value of $n$ favors the Turing pattern formation. The effect of the parameter $m$ is exactly the opposite, as its increase hinders the mechanism of pattern formation.
In Section \ref{Sec2}, we shall obtain the Turing pattern forming region in terms of three key system parameters. This will enlighten the crucial role of nonlinear diffusion to achieve pattern formation even in the case not allowed when the mechanism is driven by linear diffusion.
In Section \ref{Sec3} we shall perform the weakly nonlinear (WNL) analysis  to derive the equation
ruling the evolution of the amplitude of
the most unstable mode, both in the supercritical (Stuart-Landau equation) and the subcritical case (quintic Stuart-Landau equation). In Section 4 we shall address the process of pattern formation
in the vicinity of the Hopf bifurcation, when it is the complex Ginzburg-Landau equation to provide a spatio-temporal
modulation of the phase and amplitude of the
homogeneous oscillatory solution \cite{AK02}.

\section{Linear instabilities}\label{Sec2}
\setcounter{figure}{0}

In analogy with \cite{GLSS13,GMV08}, we rescale \eqref{or_syst} as follows:
\begin{equation}\label{syst}
\begin{split}
\frac{\partial u}{\partial t}=&\,\frac{\partial^2}{\partial x^2} u^{m+1}+\Gamma\left(a-u-\frac{4uv}{1+u^2}\right),\\
\frac{\partial v}{\partial t}=&\,c d\frac{\partial^2}{\partial x^2}v^{n+1}+\Gamma c b\left(u-\frac{uv}{1+u^2}\right),
\end{split}
\end{equation}
using $U=u,\ V=v,\ \tau=t,\ \zeta=x^*x\ $, where:
\begin{equation}\label{nodim}
x^*=\sqrt{\frac{D_u}{(m+1)u_0^m}},
\end{equation}
and the parameter $d$ has been defined as:
\begin{equation}\label{diff_n}
d=\frac{(m+1)D_vu_0^m}{(n+1)D_uv_0^m}.
\end{equation}
In what follows the system \eqref{syst} is supplemented with initial data and, given that
we are interested in self-organizing patterns, we impose
Neumann boundary conditions.

\begin{figure}[h]
\begin{center}
{\epsfxsize=2.1 in \epsfbox{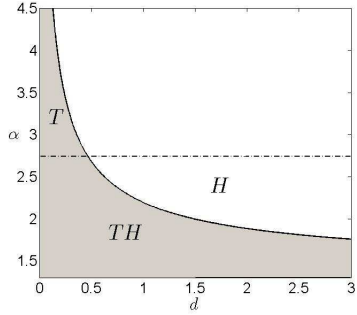}}
\end{center}
\caption{\label{f1}  The instabilities region. Here the parameters are chosen as $m=n=1$, $c=8$ and $b=1.2$. }
\end{figure}

\begin{figure}[h]
\begin{center}
\subfigure{\epsfxsize=2.1 in \epsfbox{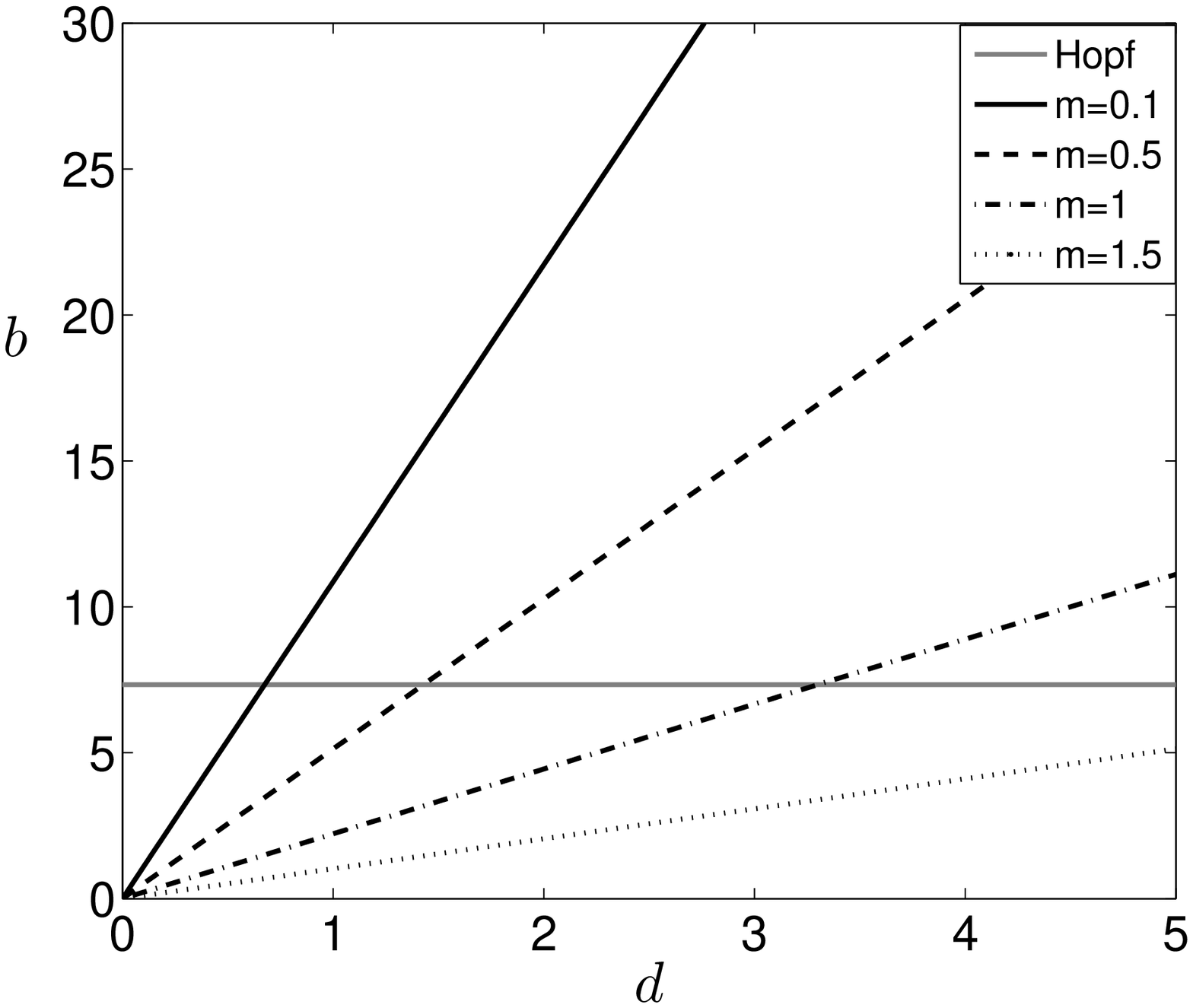}}
\subfigure{\epsfxsize=2.1 in \epsfbox{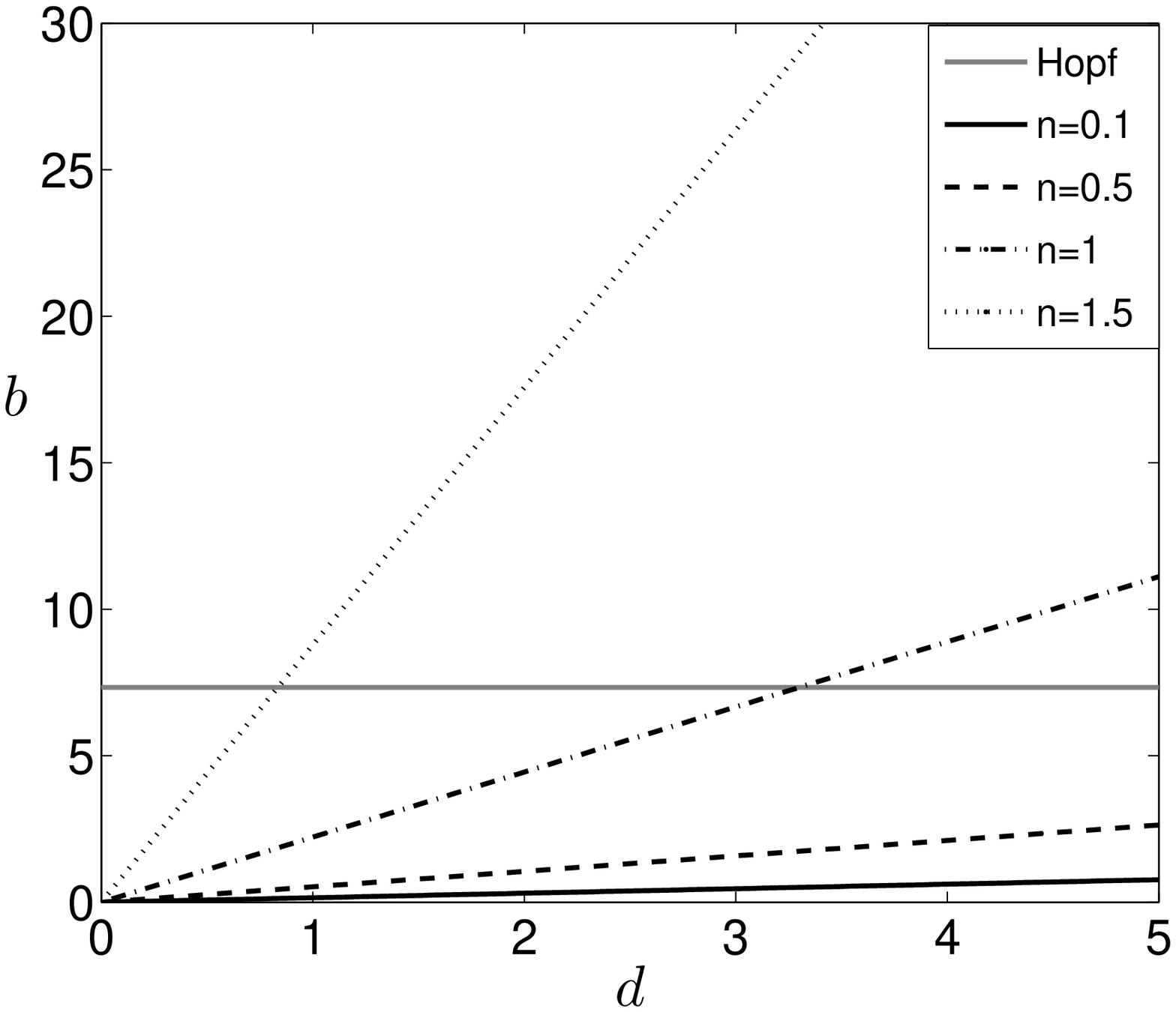}}
\end{center}
\caption{\label{f2}  Turing and Hopf instability boundaries varying $m$ and $n$. The instabilities stay below the lines.}
\end{figure}

The only homogeneous stationary state admitted by the system \eqref{syst} is
 $(\bar{u},\bar{v})\equiv (\alpha, 1+\alpha^2)$, where $\alpha=a/5$.
Carrying out the linear stability analysis, we derive the dispersion relation $\lambda^2+g(k^2)\lambda+h(k^2)=0$ which gives the growth
rate $\lambda$ as a function of the wavenumber $k$, where:
\[\begin{split}
g(k^2)=&\; k^2 \,{\rm tr}(D)-\Gamma \,{\rm tr}(K) ,\\
h(k^2)=&\;{\rm det}(D)k^4+\Gamma q k^2+\Gamma^2 {\rm det}(K),\end{split}\]
with:
\begin{equation}\label{2.2}
K=\Gamma\left(\begin{array}{cc}
\displaystyle\frac{3\alpha^2-5}{\alpha^2+1}  & -\displaystyle\frac{4\alpha}{\alpha^2+1} \\
\displaystyle\frac{2c b\alpha^2}{\alpha^2+1}  & -\displaystyle\frac{c b}{\alpha^2+1}
\end{array}\right),\qquad
D=\left(\begin{array}{cc}
(m+1)\bar{u}^m & 0\\
0 & c d(n+1)\bar{v}^n
\end{array}\right)\!,
\end{equation}
and $q=-K_{11}D_{22}-K_{22}D_{11}$.
Notice that the system \eqref{syst} is an activator-inhibitor system under the condition:
\begin{equation}\label{AIcond}
3\alpha^2 - 5>0,
\end{equation}
since $K_{11}>0, K_{22}<0, K_{12}<0$ and $K_{21}>0$ (see discussion of activator-inhibitor systems
in \cite{Mur07}).
The steady state $(\bar{u},\bar{v})$ can lose its stability both via Hopf and Turing bifurcation.
Oscillatory instability occurs when $g(k^2)=0$ and $h(k^2)>0$.
The minimum values of $b$ and $k$ for which $g(k^2)=0$ are:
\begin{equation}\label{hopf}
b_H=\frac{3\alpha^2-5}{c\alpha}\, \qquad  k=0,
\end{equation}
and for $b<b_H$ a spatially homogeneous oscillatory mode emerges. Notice that condition \eqref{AIcond} assures $b_H>0$.
The neutral stability Turing boundary corresponds to $h(k^2)=0$,
which has a single minimum $(k_c^2, b_c)$ attained when:
\begin{equation}\label{kc2}
k_c^2=-\frac{\Gamma q}{2\, \rm{det} (D)} \, ,
\end{equation}
which requires $q<0$. Therefore a necessary condition for Turing instability is given by:
\begin{equation}\label{q_neg}
b<\bar{b}=d\frac{(n+1)}{(m+1)} \frac{(3\alpha^2-5)(1+\alpha^2)^n}{\alpha^{m+1}}.
\end{equation}
The value $\bar{b}$ is non-negative under the condition \eqref{AIcond}. Moreover it can be straightforwardly proved
that $\bar{b}$ is a decreasing function of $m$ and an increasing function of $n$, which means that larger values of $n$
facilitates Turing instability occurring for any values of $d$. However, when $n=0$, once fixed $m\geq 0$, in order to satisfy the condition \eqref{q_neg},
the value of $d$ should be sufficiently large, i.e. the diffusion of the inhibitor should be greater than
that of the activator.
Substituting the expression in \eqref{kc2} for the most unstable mode in $h(k_c^2)=0$,
the Turing bifurcation value $b=b_c$ is obtained by imposing:
\begin{equation}\label{tur2}
q^2 -4\ {\rm det}(D){\rm det}(K)=0,
\end{equation}
under the condition $q<0$. Introducing $b=\bar{b}-\xi$, with $\xi>0$, in \eqref{tur2} one gets:
\begin{equation}\label{xi_tur}
\begin{split}
(m+1)\frac{\alpha^{m+1}}{(1+\alpha^2)^2}\,\xi^2+20(n+1)d(\alpha^2+1)^{n-1}\xi\\
-20(n+1)^2d^2(\alpha^2+1)^{2n-1}&\frac{3\alpha^2-5}{\alpha^{m+1}}=0,
\end{split}
\end{equation}
whose positive root:
\begin{equation}\label{xi_p}
\begin{split}
\xi=\xi^+=2d\frac{(n+1)(\alpha^2+1)^n}{(m+1)\alpha^{m+1}}\left(\sqrt{5(3\alpha^4m+8\alpha^4-2\alpha^2m+8\alpha^2-5m)}\right.\\
\left.-5(\alpha^2+1)\right)&
\end{split}
\end{equation}  (this choice
guarantees the condition $q<0$) gives the critical value of the parameter $b$:
\begin{equation}\label{bicritic}
b_c=\bar{b}-\xi^+=d \frac{(1+\alpha^2)^n}{\alpha^{m+1}}\frac{n+1}{m+1}\left(13\alpha^2+5-4\alpha\sqrt{10(1+\alpha^2)}\right) \;,
\end{equation}
which is nonnegative under the condition \eqref{AIcond}.
In Fig.\ref{f1} we show the instability regions in the parameter space $(d,\alpha)$: the Turing instability region T, the Hopf instability
region H and the region TH where a competition
between the two instabilities occurs. In TH which one would develop,
depends on the locations
of the respective instability boundaries: as $b$ decreases, if
$b_c > b_H$, Turing instability occurs prior to the oscillatory
instability and the Turing structures form.
Imposing $b_c \geq b_H$ leads to the following inequality:
\begin{equation}\label{TH}
d\geq \bar{s}=\frac{(m+1)\alpha^m(3\alpha^2-5)}{(n+1)c(1+\alpha^2)^n(13\alpha^2+5-4\alpha\sqrt{10(1+\alpha^2)})},
\end{equation}
where the value $\bar{s}$ is nonnegative under the condition \eqref{AIcond}. Moreover, it can be easily proved that $\bar{s}$
is an increasing function with respect to $m$ and a decreasing function with respect to $n$, which means that larger values of $n$
favors Turing instability and the formation of the relative pattern,
also when the parameter $d$ is small.
The effect of the parameter $m$ is opposite.
This is also evident in Fig.\ref{f2}, where the Turing and the Hopf instabilities boundaries are drawn with respect to the parameter $d$ varying $m$ and $n$.

\section{WNL analysis and pattern formation}\label{Sec3}
We use the method of multiple scales to determine the amplitude equation of the pattern close to the instability threshold.
Introducing the control parameter $\varepsilon$, which represents the dimensionless distance
from the critical value and is therefore defined as $\varepsilon^2=(b-b_c)/b_c$,
the characteristic slow temporal scale $T=\varepsilon^2 t$ can be easily obtained via linear analysis (see \cite{GLS12}).
Let us recast the original system \eqref{syst} in the following form:
\begin{equation}\label{3.1}
\partial_t \textbf{w}= \mathcal{L}^{b} \textbf{w}+
\mathcal{NL}^{b} \textbf{w},\qquad
\textbf{w}\equiv\left(\begin{array}{c}{u-\bar{u}}\\
{v-\bar{v}}\end{array}\right) \; ,
\end{equation}
where the linear operator $\mathcal{L}^{b}=\Gamma \, K^b + D \nabla^2$ and the nonlinear operator
$\mathcal{NL}^{b}$ contains the remaining terms.
The matrix $K^b$ and $D$ are given in \eqref{2.2}, we made explicit the
dependence on the bifurcation parameter ${b}$ just for notational convenience.

Passing to the asymptotic analysis, we expand  $b$ and
$\textbf{w}$ as:
\begin{eqnarray}\label{3.3}
b &=& b_c+\varepsilon^2 b^{(2)}+\varepsilon^4 b^{(4)}+\dots ,
\\
\textbf{w}&=&\varepsilon \, \textbf{w}_1 +\varepsilon^2\,
 \textbf{w}_2+\varepsilon^3\,
 \textbf{w}_3+\dots , \label{3.3bis}\\
 t&=&t+\varepsilon^2 T_2+\varepsilon^4 T_4+\dots,  \label{3.3ter}
\end{eqnarray}
where the coefficients $b^{(i)}$ are negative.
Substituting \eqref{3.3}-–\eqref{3.3ter} into the full system \eqref{3.1}, the following sequence of linear equations for $\textbf{w}_i$ is
obtained:
\\
$\ \,O(\varepsilon):$
\begin{equation}\label{sequence_1}
\mathcal{L}^{b_c} {\bf w}_1=\mathbf{0},
\end{equation}
$\ \,O(\varepsilon^2):$
\begin{equation}\label{sequence_2}
\mathcal{L}^{b_c} {\bf w}_2=\mathbf{F},
\end{equation}
$\ \,O(\varepsilon^3):$
\begin{equation}\label{sequence_3}
\mathcal{L}^{b_c} {\bf w}_3=\mathbf{G},
\end{equation}
where:
\[
\mathbf{F}=\frac{\partial {\bf
w}_1}{\partial T_1}-D^{(1)}\nabla^2\left(\begin{array}{c} u_1^2\\v_1^2\end{array}\right)
+\frac{\alpha((\alpha^2-3)u_1-(\alpha-1)v_1)}{(\alpha^2+1)^2}\mathfrak{u}_1,
\]
\[
\begin{split}
\mathbf{G}=&\,\frac{\partial {\bf
w}_1}{\partial T_2}-D^{(2)}\nabla^2
\left(\begin{array}{c}
u_1^3\\
v_1^3
\end{array}
\right)
-
2D^{(1)}\nabla^2
\left(\begin{array}{c}
u_1u_2\\
v_1v_2
\end{array}
\right)
-\frac{c b^{(2)}\alpha(2\alpha u_1-v_1)}{\alpha^2+1}\mathfrak{u}_2
\\
-&\,
\frac{(\alpha^4-6\alpha^2+1)u_1^3-\alpha(\alpha^2-3)u_1v_1+(\alpha^4-1)(u_1v_2+u_2v_1)}{(\alpha^2+1)^3}\mathfrak{u}_1
\\
-&\,
\frac{2\alpha(-\alpha^4+2\alpha^2+3)u_1}{(\alpha^2+1)^3}\mathfrak{u}_1
\end{split}
\]
and:
\begin{equation}\nonumber
\mathfrak{u}_1=\Gamma\left(\begin{array}{c} 4\\c b_c\end{array}\right),\qquad \qquad \mathfrak{u}_2=\Gamma\left(\begin{array}{c} 0\\1\end{array}\right),
\end{equation}
\begin{equation}\nonumber
D^{(1)}=\left(
\begin{array}{cc}
\displaystyle\frac{m(m+1)}{2}
\bar{u}^{m-1} & 0\\
0 & cd\displaystyle\frac{n(n+1)}{2}\bar{v}^{n-1}
\end{array}
\right),
\end{equation}
\begin{equation}\nonumber
D^{(2)}=\left(
\begin{array}{cc}
\displaystyle\frac{m(m^2-1)}{6}
\bar{u}^{m-2} & 0\\
0 & \displaystyle\frac{n(n^2-1)}{6}\bar{v}^{n-2}
\end{array}
\right).
\end{equation}
At the lowest order in $\varepsilon$ we recover the linear problem
$\mathcal{L}^{b_c} {\bf w}_1=\mathbf{0}$ whose solution,
satisfying the Neumann boundary conditions, is given by:
\begin{equation} \label{3.5}
{\bf w}_1=A(T) \bfrho \, \cos(\bar{k}_c x) \; , \qquad \mbox{with}
\quad \bfrho \in \mbox{Ker}( K^{b_c}-\bar{k}_c^2D)\; .
\end{equation}
In the above equation we have denoted with $\bar{k}_c$  the first admissible unstable mode,
while $A(T)$ is the amplitude of the pattern and it is
still arbitrary at this level. The vector $\bfrho$ is defined up to a
constant and we shall make the normalization in the following way:
\begin{equation}\label{emme}
\bfrho =  \left(\begin{array}{c} 1 \\M \end{array}\right) \, ,
\qquad \mbox{with} \quad
 M\equiv\frac{-D_{21}k_c^2+\Gamma K^{b_c}_{21}}{D_{22}k_c^2-\Gamma K^{b_c}_{22}},
\end{equation}
where $D_{ij}, K^{b_c}_{ij}$ are the $i,j$-entries of
the matrices $D$ and $K^{b_c}$.

\begin{figure}[h]
\begin{center}
{\epsfxsize=2.1 in \epsfbox{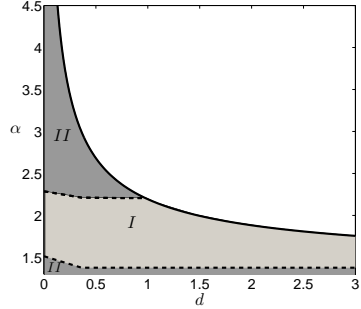}}
\end{center}
\caption{\label{SubSup}  The Turing region: subcritical and supercritical. The parameters are chosen as $m=n=1$, $c=8$ and $b=1.2$.}
\end{figure}

\begin{figure}[h]
\begin{center}
\subfigure{\epsfxsize=2.3 in \epsfbox{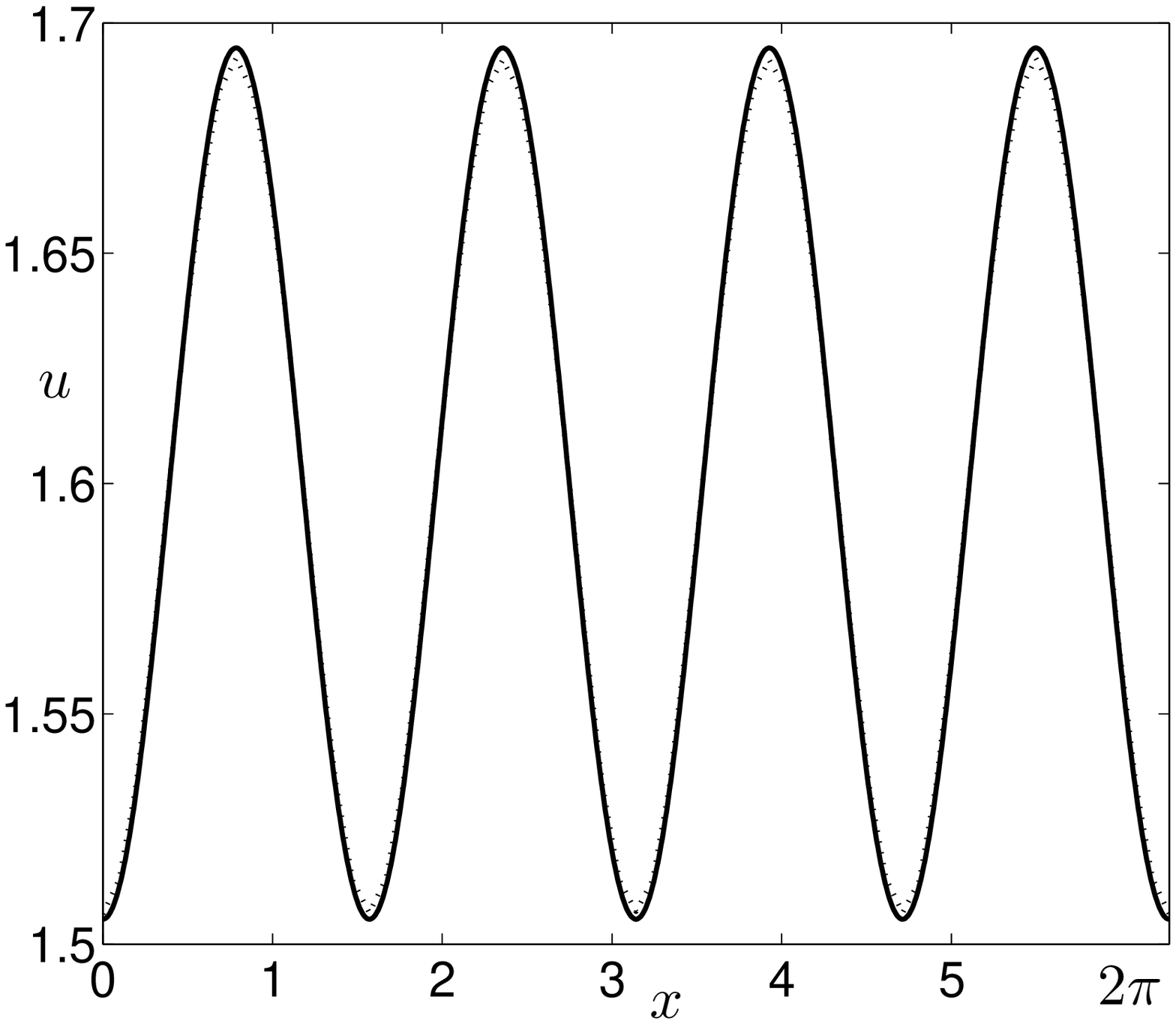}}
\end{center}
\caption{\label{super} Comparison between the WNL solution (solid line) and the numerical solution of \eqref{syst} (dotted line) in the supercritical case. The parameters are chosen as $m=n=1$, $\Gamma=140$, $\alpha=1.6$ $c=40$, $d=1.5$, $b_H=0.0419<b=b_c(1-0.1^2)<b_c\sim 0.1959$ and $k_c=4$.}
\end{figure}

Once substituted in \eqref{sequence_2} the first order solution ${\bf w}_1$, the vector ${\bf F}$ is orthogonal
to the kernel of the adjoint of $ \mathcal{L}^{b_c}$ and the equation \eqref{sequence_2} can be solved right away.
This is not the case for Eq.\eqref{sequence_3}. In fact the vector $\mathbf{G}$ has the following
expression:
\begin{equation}\label{G}
{\bf G}=\left(\displaystyle\frac{d A}{d
T_2}\bfrho+A {\bf G}_1^{(1)}+A^3 {\bf G}_1^{(3)}
\right)\cos(\bar{k}_c x)+{\bf G}^* ,
\end{equation}
where ${\bf G}^{*}$ contains automatically
orthogonal terms and ${\bf G}_1^{(j)}, j = 1,3$ have a cumbersome expression here not reported.
The solvability condition for the equation \eqref{sequence_3} gives the following Stuart-Landau equation (SLE)
for the amplitude $A(T)$:
\begin{equation}\label{3.11}
\frac{d A}{d T}= \sigma A -L A^3,
\end{equation}
where the coefficients $\sigma$ and $L$ are given as follows:
\begin{equation}\label{sigmaL}\nonumber
\sigma=-\frac{<{\bf G}_1^{(1)}, \bfpsi>}{<\bfrho,
\bfpsi>},\qquad  L=\frac{<{\bf G}_1^{(3)}, \bfpsi>}{<\bfrho,
\bfpsi>},\quad {\rm and\ }{\bfpsi} \in
{\rm Ker}\left(K^{b^c} -\bar{k}_c^2D
\right)^\dag.
\end{equation}
%
%
Since the growth rate coefficient $\sigma$ is always
positive, the dynamics of the SLE \eqref{3.11}
can be divided into two qualitatively different cases depending on
the sign of the Landau constant $L$: the supercritical case, when
$L$ is positive, and the subcritical case, when $L$ is negative. In Fig.\ref{SubSup}, the curves across which $L$ changes its sign
are drawn in the space $(d, \alpha)$ and the
pattern-forming region is divided in one supercritical region (I) and two subcritical regions (II).
In the supercritical case $A_\infty=\sqrt{{\sigma}/{L}}$ is the stable equilibrium
solution of the amplitude equation \eqref{3.11} and it corresponds to the
asymptotic value of the amplitude $A$ of the pattern.
In Fig.~\ref{super} we show the comparison between the solution predicted by
the WNL analysis up to the $O(\varepsilon^2)$ and the stationary state (reached starting from a random perturbation
of the constant state) computed solving numerically the full system \eqref{syst}.
In all the performed tests, we have checked that the distance between the WNL approximated solution and
the numerical solution of the system \eqref{syst} is $O(\varepsilon^3)$ in
$L^1$ norm.

\begin{figure}[h]
\begin{center}
\subfigure{\epsfxsize=2.1 in \epsfbox{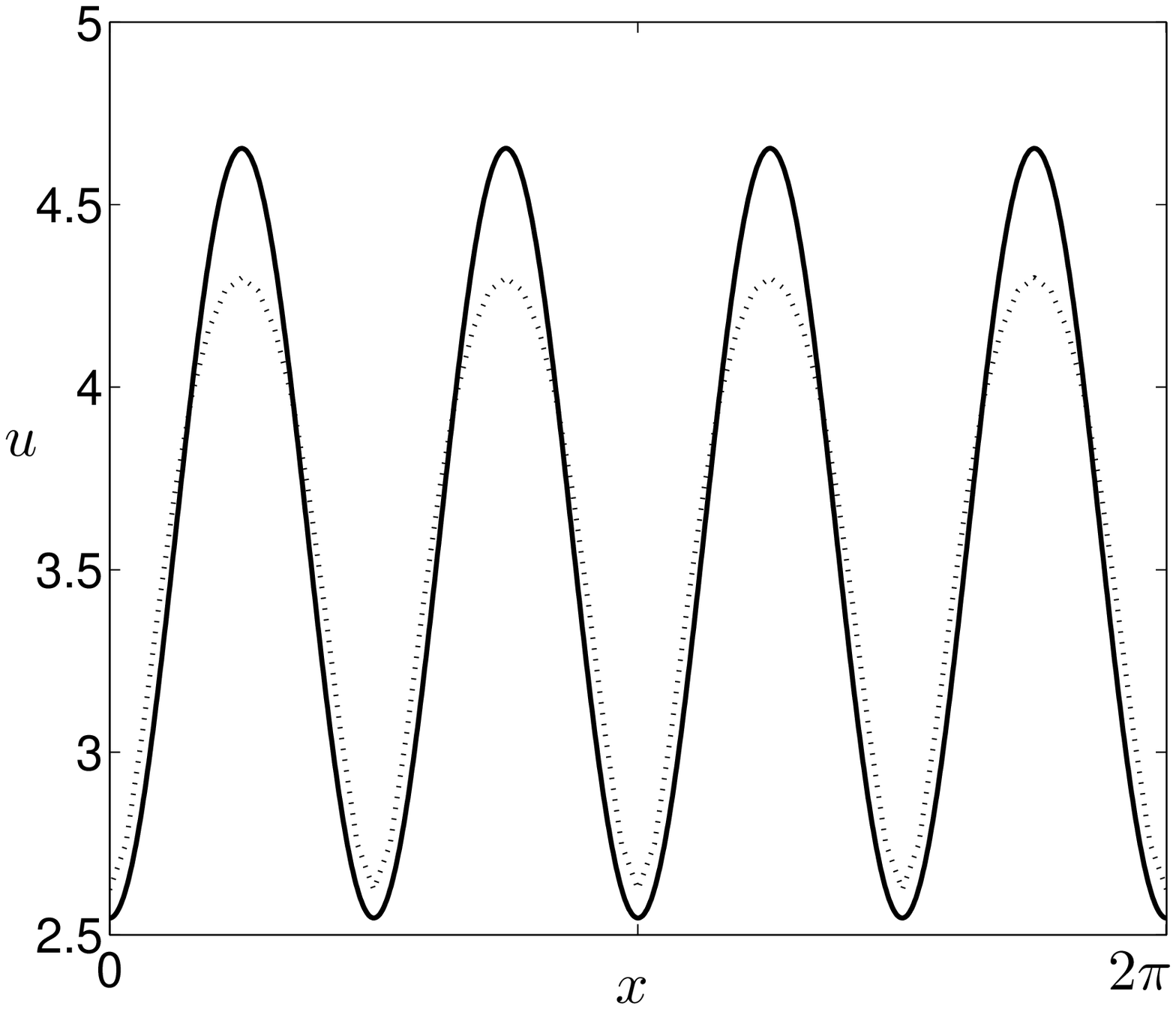}}
\subfigure{\epsfxsize=2.1 in \epsfbox{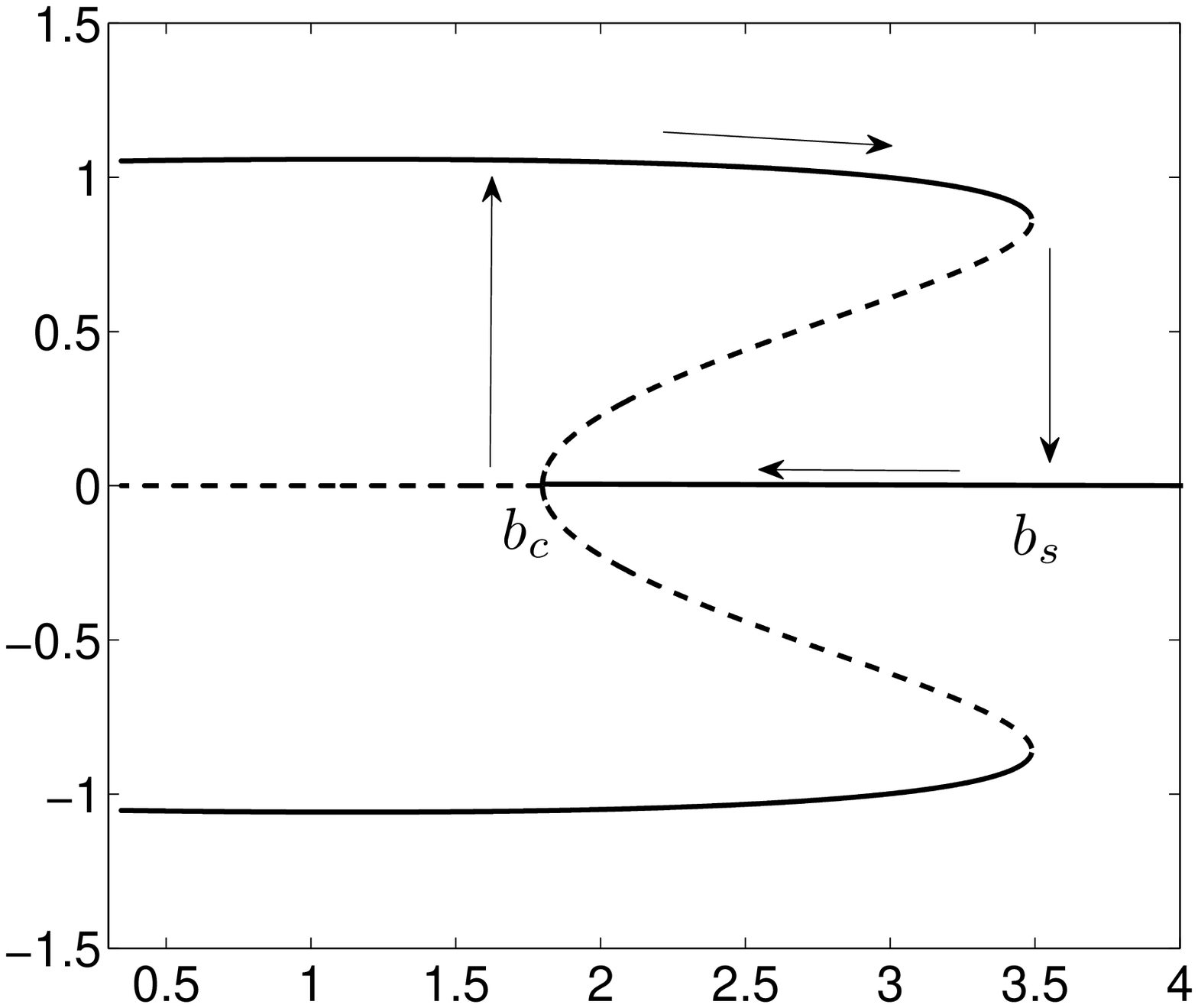}}
\end{center}
\caption{\label{sub}  (a) Comparison between the WNL solution (solid line) and the numerical solution of \eqref{syst} (dotted line) in the subcritical case.
(b) The bifurcation diagram in the subcritical case. The parameters are chosen as $m=n=1$, $\Gamma=100$, $\alpha=3.6$ $c=40$, $d=0.5$, $b_H\sim0.2353<b=b_c(1-0.1^2)<b_c\sim 1.7991$ and $k_c=4$ .}
\end{figure}

In the subcritical regions indicated with II in Fig.\ref{SubSup},
the Landau coefficient $L$ has a
negative value and the equation \eqref{3.11} is not able to capture
the amplitude of the pattern. In this
case to predict the amplitude of the pattern, one
needs to push the WNL expansion at a higher order
(for a general discussion on the relevance of the higher order
amplitude expansions in the study of subcritical bifurcations, see
the recent \cite{BMS09} and references therein).
Performing the WNL up to $O(\varepsilon^5)$
we obtain the following quintic SLE
for the amplitude $A$:
\begin{equation}\label{quintic_SL}
\frac{d A}{d
T_2}=\bar{\sigma}A-\bar{L}A^3+\bar{Q}A^5\, .
\end{equation}
Here we skip the details of the analysis, but we want to stress that the coefficients
$\bar{\sigma}$ and $\bar{L}$ are $O(\varepsilon^2)$ perturbation of the coefficients $\sigma$
and $L$ of the SLE \eqref{3.11}, and the coefficient $\bar{Q}$ is $O(\varepsilon^2)$.
The predicted amplitude is $O(\varepsilon^{-1})$, and therefore the corresponding
emerging pattern is an $O(1)$ perturbation of the equilibrium,
which contradicts the basic assumption of the perturbation scheme \eqref{3.3bis}.
In the subcritical case, when the growth rate coefficient
$\bar{\sigma}>0$, the Landau coefficient $\bar{L}<0$ and
$\bar{Q}<0$, one should therefore expect
quantitative discrepancies between the predicted solution of the WNL analysis and the
numerical solution of the full system. Nevertheless in our numerical tests we have found a
qualitatively good agreement, see for example Fig.\ref{sub}(a).
Moreover, the bifurcation diagram in Fig.\ref{sub}(b) constructed using the amplitude equation \eqref{quintic_SL} is able to predict
very well phenomena like bistability and hysteresis cycle shown also by the full system,
see \cite{GLS09}.
\section{Oscillating pattern at the Hopf bifurcation}
\begin{figure}[h]
\begin{center}
\subfigure{\epsfxsize=2.3 in \epsfbox{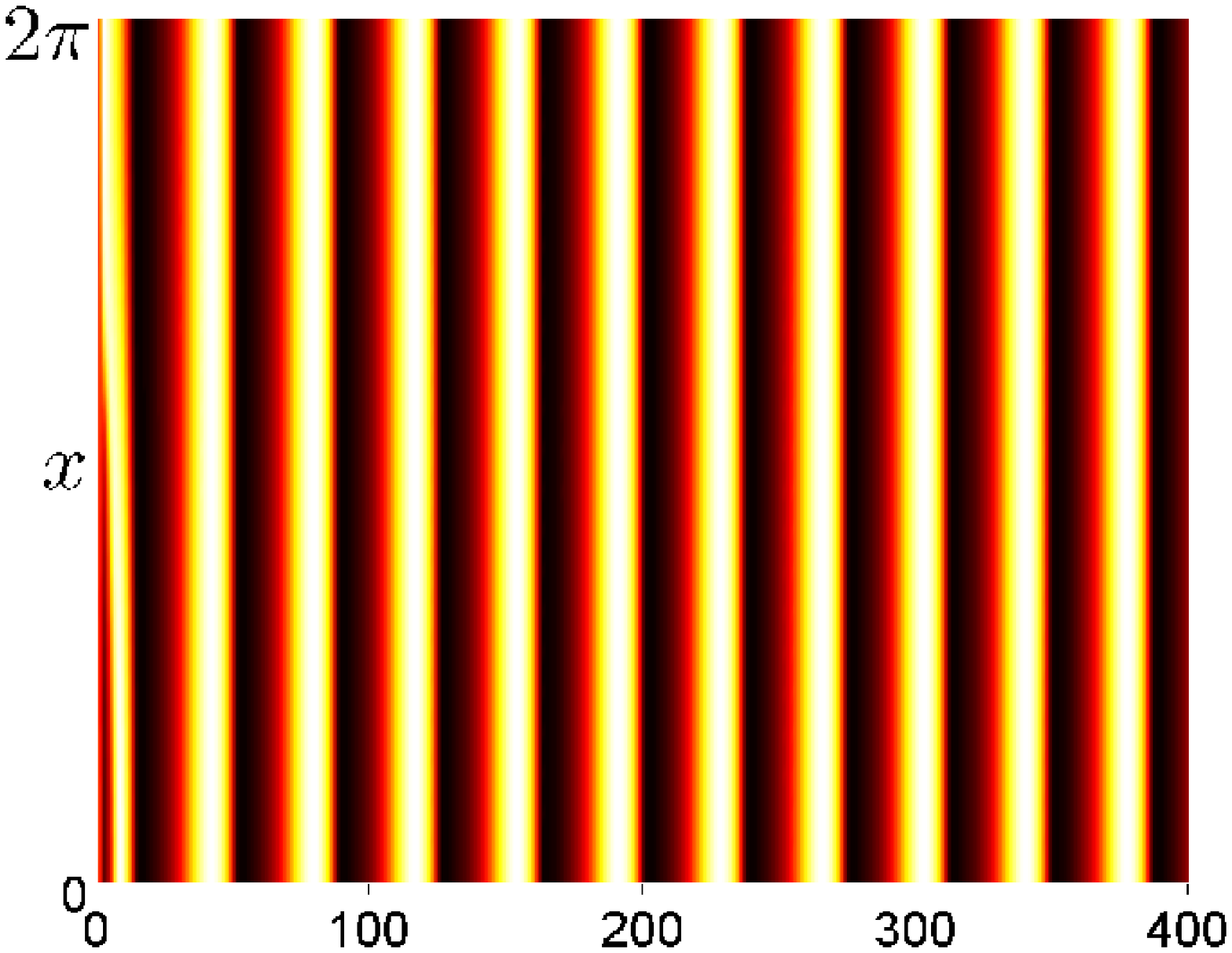}}
\subfigure{\epsfxsize=2.3 in \epsfbox{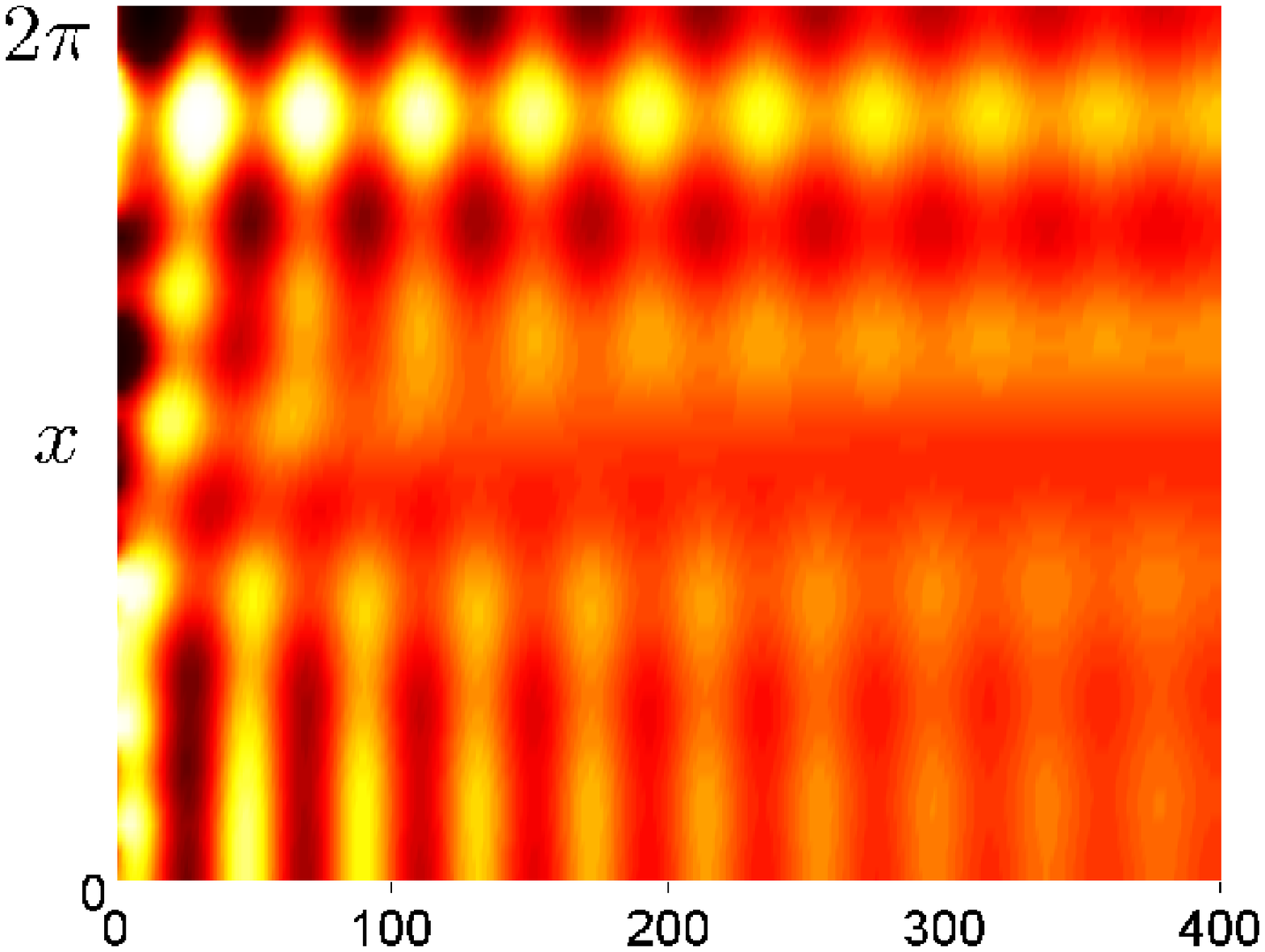}}
\end{center}
\caption{\label{oscill} Oscillating patterns. (a) Stable periodic solution arising from the oscillation of the equilibrium $(\bar{u}, \bar{v})$. The parameters are chosen as $m=n=1$, $\Gamma=140$, $\alpha=1.6$ $c=2$, $d=1.5$, $b_c\sim 0.1959<b=b_H(1-0.1^2)<b_H\sim0.8375$. (b) A Turing-type pattern oscillates next to the codimension 2 Turing-Hopf bifurcation point. The parameters are chosen as $m=n=1$, $\Gamma=140$, $\alpha=1.58$ $c=40$, $d=0.3206$, $b_c\sim 0.0372$ $b_H\sim 0.0394$, $b=0.03787$}
\end{figure}

In the Hopf instability region, labeled with H in Fig.\ref{f1}, the solution of the system \eqref{syst} has a pure oscillating dynamical behavior,
see Fig.\ref{oscill}(a) where the homogeneous state $(\bar{u}, \bar{v})$ destabilizes
and a stable periodic solution emerges.
When the values of the Turing bifurcation point $b_c$ and the Hopf bifurcation point $b_H$ are rather close, our numerical investigations
have also shown that, even though the parameter $b$ is chosen into the Hopf instability region, the proximity to the Turing instability region influence
the emerging solution: there is a transient in which a Turing structure oscillates and the corresponding solution in the time-space plane is given in Fig.\ref{oscill}(b).
In the direct numerical simulations of the full system \eqref{syst}, the integrator must use a step size sufficiently small to follow all the oscillations.
The complex Ginzburg-Landau equation
gives a universal description of reaction-diffusion systems in the neighborhood of the Hopf bifurcation.
Using the same asymptotic expansion as in \eqref{3.3}, where the bifurcation value is now $b_H$,
and taking into account also the slow spatial modulation $X$ (whose characteristic length scale is $O(\varepsilon^{-1}$)),
at the lowest order $\varepsilon$ we recover the linear problem $\mathcal{L}^{b_H}
{\bf w}_1=\mathbf{0}$, whose solution is:
\begin{equation} \label{sol1H}
{\bf w}_1=\mathcal{A}(X,T)\bfthe e^{ih_c t}+ \bar{\mathcal{A}}(X,T)\overline{\bfthe} e^{-ih_c t}\; ,
\end{equation}
where $h_c=\sqrt{{\rm det}(K^{b_H})}$ and the vectors $\bfthe$ and $\overline{\bfthe}$ satisfy $K^{b_H}\bfthe=ih_c\bfthe$, and $\overline{\bfthe} K^{b_H}=ih_c\overline{\bfthe}$.
%
%
Pushing the asymptotic analysis up to $O(\varepsilon^3)$
(the details are not reproduced here as they follow the same steps as in Section\ref{Sec3}),
we find the following complex Ginzburg-Landau equation (CGLE) for the amplitude $\mathcal{A}$:
\begin{equation} \label{complexGL}
\frac{\partial \mathcal{A}}{\partial T}=\delta \frac{\partial^2 \mathcal{A}}{\partial X^2}+\chi \mathcal{A}+\eta |\mathcal{A}|^2\mathcal{A},
\end{equation}
where the coefficients $\eta$ and $\delta$ are complex and the coefficient $\chi$ is real.
The amplitude $\mathcal{A}$ describes the modulation of local oscillations having frequency $h_c$ and the
fact that the fundamental phase $e^{i h_c t}$has been scaled out in the CGLE has enormous numerical
advantages.

\section{Conclusions and open problems}
In the present paper we have examined the Turing mechanism induced by a
nonlinear density-dependent diffusion in the Lengyel-Epstein
system. We have shown that the presence of nonlinear diffusion favors the Turing instability (which competes with the Hopf instability)
and the formation of Turing structure also when the diffusion coefficient of the
activator exceeds that one of the inhibitor. Through a WNL analysis, we have derived the equations
which rule the amplitude the pattern, both in the supercritical and subcritical bifurcation case, identifying in the parameters space the
supercritical and the subcritical regions. All the numerical tests we have run are in good agreement with the prediction of the WNL analysis.
We have also numerically investigated the oscillating pattern arising in the Hopf instability region and we have computed
the CGLE as it describes the slow spatio-temporal
modulation of the amplitude of the
homogeneous oscillatory solution.
Some other aspects of the problem could be examined.
In a 2D domain new pattern forming phenomena occur, as degeneracy leads to more complex structures,
predictable via the WNL \cite{GLS13}.
Moreover, the analytic solutions of the CGLE
can be obtained in some special cases, e.g. plane wave solutions. Even though they are the simplest propagating
structures supported by the CGLE, they have a fundamental role as their criteria of stability are necessary conditions for spiral wave instability.
Finally, we could explore the spatio-temporal chaos in the starting from the numerical investigation of the spatio-temporal chaos in the CGLE \cite{BC12,BP06}.


\bibliographystyle{plain}
\bibliography{pattern}

\end{document}